\documentclass{article}
\usepackage{spconf,amsmath,graphicx,amssymb}
\usepackage{cite, url, multirow, bm, subfigure}

\title{Neural Diarization with Non-autoregressive Intermediate Attractors}
\name{
Yusuke Fujita$^{\star}$,
Tatsuya Komatsu$^{\star}$,
Robin Scheibler$^{\star}$,
Yusuke Kida$^{\star}$,
Tetsuji Ogawa$^{\dagger}$
}
\address{$^{\star}$LINE Corporation, Japan \quad $^{\dagger}$Waseda University, Japan}

\def\*#1{\bm{#1}}

\begin{document}
\ninept
\maketitle
\begin{abstract}
End-to-end neural diarization (EEND) with encoder-decoder-based attractors (EDA) is a promising method to handle the whole speaker diarization problem simultaneously with a single neural network.
While the EEND model can produce all frame-level speaker labels simultaneously, it disregards output label dependency.
In this work, we propose a novel EEND model that introduces the label dependency between frames.
The proposed method generates non-autoregressive intermediate attractors to produce speaker labels at the lower layers and conditions the subsequent layers with these labels.
While the proposed model works in a non-autoregressive manner, the speaker labels are refined by referring to the whole sequence of intermediate labels.
The experiments with the two-speaker CALLHOME dataset show that the intermediate labels with the proposed non-autoregressive intermediate attractors boost the diarization performance.
The proposed method with the deeper network benefits more from the intermediate labels, resulting in better performance and training throughput than EEND-EDA.

\end{abstract}
\section{Introduction}
\label{sec:intro}


Speaker diarization is the task of detecting multi-speaker speech activity in audio recordings.
It has been actively studied as an essential component for conversational speech understanding \cite{Tranter2006, Anguera2007, park2022review}.
The task has been evaluated in telephone conversations (CALLHOME \cite{callhome}), meetings (ICSI \cite{Janin03, etin2006OverlapIM}, AMI \cite{Renals2008}), web videos (VoxConverse \cite{chung20_interspeech}) and various hard scenarios (DIHARD Challenges \cite{Sell2018dihard,Ryant2019,ryant21_interspeech}).

A standard approach to speaker diarization is speaker embedding clustering \cite{Shum2013, Sell2014, Romero2017, Wang2018LSTM}, which first extracts speaker-discriminative embeddings like x-vectors \cite{Snyder2018} and d-vectors \cite{Wan2018} for fixed-length speech segments, and then merges homogeneous segments to the same speaker by applying a clustering algorithm such as spectral clustering \cite{Wang2018LSTM} and agglomerative hierarchical clustering \cite{Sell2014, Romero2017}.
For speaker embedding clustering, voice activity detection should be done in advance to determine speech/non-speech boundaries, and overlapping speech segments have to be eliminated with some pre- or post-processing methods\cite{Landini21}.

As an alternative to speaker embedding clustering, end-to-end neural diarization (EEND) \cite{Fujita2019E2EDiarization} has been proposed.
EEND learns a neural network to directly produce {\it full} speaker diarization labels containing speech/non-speech boundaries and overlapping speech segments, as well as speaker assignments for the detected speech segments.
Since the speaker diarization labels have permutation ambiguity due to the arbitrary order of speaker indices, the EEND network is trained with permutation-invariant training objectives.
Various network architectures have been investigated for EEND including Transformer \cite{Fujita2019ASRU}, time-dilated convolutional neural network \cite{Maiti21ICASSP}, and Conformer \cite{liu21j_interspeech}.
Encoder-decoder-based attractor (EDA) \cite{Horiguchi2020, Horiguchi2022_taslp} is a promising network architecture for EEND, which first generates an attractor vector for each speaker from an embedding sequence, and then generates the speaker's activity by measuring the similarity between the embedding sequence and the attractor vector.

Although the EEND-EDA model is found to be effective compared with speaker embedding clustering, there are two points of improvement.
First, the model assumes conditional independence between frame-level speaker labels, which limits their performance.
The EEND-EDA model runs in a non-autoregressive manner and produces all frame-level speaker labels in parallel.
However, such non-autoregressive models disregard the potential benefit of the output label dependency between frames.
Second, the long-short term memory (LSTM)-based encoder in EDA receives the frame-level embeddings recursively.
The well-known vanishing gradient problem of LSTM hinders the optimization of the lower layers.

In related fields, researchers have studied the use of ``intermediate'' labels to relax the conditional independence assumption in such non-autoregressive models.
For non-autoregressive automatic speech recognition (ASR) based on connectionist temporal classification (CTC), Intermediate CTC \cite{lee21_icassp} is proposed to introduce auxiliary tasks of predicting labels inside the network by inserting the same CTC losses to the intermediate predictions.
Self-conditioned CTC \cite{nozaki21_interspeech} further utilizes the intermediate labels at the lower layer as conditions for enhancing the predictions at the upper layers.
This self-conditioning technique achieves the best performance among the non-autoregressive ASR systems \cite{Higuchi21_asru}.
For EEND, a similar intermediate speaker label prediction technique \cite{Yu22_icassp} is proposed, which uses the same permutation-invariant training objectives for the intermediate speaker labels.
However, the intermediate speaker labels are not utilized for the self-conditioning features.
These prior studies motivate us to introduce the self-conditioning technique in EEND models.

In this paper, we propose a novel network architecture for EEND that uses intermediate speaker labels to condition the subsequent network layers.
For producing the intermediate speaker labels, the proposed method extracts intermediate attractors at every encoder layer.
The auxiliary permutation-invariant training losses are introduced for optimizing the intermediate labels.
For conditioning the subsequent network layers, the proposed method adds the weighted intermediate attractors to the frame-level embeddings. While the proposed network still works in a non-autoregressive manner, the speaker labels are iteratively refined by referring to the whole sequence of intermediate speaker labels.
For the vanishing gradient problem in the EDA's LSTM, we adopt the attention mechanism \cite{Vaswani2017}.
Unlike LSTM, it does not suffer from vanishing gradients, thus facilitating optimization of the lower layers.
Another advantage of the attention mechanism is that it is non-autoregressive, so training throughput is much higher than LSTM when interacting with the intermediate training objectives.
The experimental results with the two-speaker CALLHOME dataset show that the intermediate labels with the proposed non-autoregressive intermediate attractors boost the diarization performance while the original EDA cannot get benefit from the intermediate labels.
The proposed method with the deeper network benefits more from the intermediate labels, resulting in better performance and training throughput than EEND-EDA.

\section{Method}
\label{sed:proposed}
This section briefly introduces EEND-EDA as our baseline system, followed by our proposed method: intermediate attractors.

\subsection{End-to-end Neural Diarization}
\label{sed:eend}
EEND formulates the speaker diarization problem as a frame-wise multi-label classification task \cite{Fujita2019E2EDiarization}.
In this paper, we denote $\*X \in \mathbb{R}^{D\times T}$ as a $T$-length sequence of $D$-dimensional audio features.
A neural network accepts $\*X$ and produces the same-length sequence of speaker label posteriors $\*Y \in [0,1]^{C\times T}$, where $C$ is the number of speakers and $[\*Y]_{c,t}$ is the probability that $c$-th speaker is speaking at time $t$. 
The network is trained to minimize the binary cross-entropy loss between the ground-truth speaker labels $\*Y^* \in \{0,1\}^{C\times T}$ and the estimated label posteriors $\*Y$:
\begin{align}
    \mathcal{L}_\mathsf{PIT}(\*Y^*, \*Y) = \frac{1}{CT} \min_{\phi \in \mathcal{P}(C)} \sum_{c=1}^C \sum_{t=1}^T \mathsf{BCE}([\*Y^*]_{\phi_c,t}, [\*Y]_{c,t}), \label{eq:pit}
\end{align}
where $\mathsf{BCE}(y^*,y) = - y^*\log y - (1-y^*)\log(1-y) $, $\mathcal{P}(C)$ is the set of all permutations of a sequence $\{1,\dots,C\}$.
This permutation-invariant training scheme \cite{Yu2017,Hershey2016} correctly handles the label ambiguity caused by the arbitrary order of speaker indices.

\subsection{EEND with Encoder-decoder-based Attractors}
\label{sed:eda}
The neural network for EEND adopted in \cite{Fujita2019ASRU, Horiguchi2020} comprises of a stack of Transformer encoders:
\begin{align}
    \*E_{l} = \mathsf{EncoderLayer}_l(\*E_{l-1}) \in \mathbb{R}^{D\times T} \qquad (1 \le l \le L), \label{eq:enc}
\end{align}
where $L$ is the number of Transformer layers and $\*E_0 = \*X$ \footnote{In this paper, we assume $\*X$ is subsampled beforehand.}.
Whereas the vanilla EEND \cite{Fujita2019E2EDiarization, Fujita2019ASRU} simply transforms $\*E_L$ to $\*Y$ by a linear layer followed by a sigmoid function, EDA \cite{Horiguchi2020} first generates speaker-wise attractor vectors $\*A = [\*a_1, \dots, \*a_C] \in \mathbb{R}^{D\times C}$:
\begin{align}
  \*A = \mathsf{EDA}(\*E_L). \label{eq:eda}
\end{align}
Here, this EDA function is implemented using long short-term memory (LSTM) layers:
\begin{align}
    (\*h_t, \*c_t) = \mathsf{LSTM}_\mathsf{enc}(\*h_{t-1}, \*c_{t-1}, [\*E_L]_{:,t}) \quad (1 \le t \le T), \label{eq:lstm_enc}\\
    (\*a_c, \*d_{c}) = \mathsf{LSTM}_\mathsf{dec}(\*a_{c-1}, \*d_{c-1}, \mathbf{0}) \quad (1 \le c \le C), \label{eq:lstm_dec}
\end{align}
where $\mathsf{LSTM}_\mathsf{enc}()$ is an unidirectional LSTM layer that sequentially reads an embedding vector for time $t$, $[\*E_L]_{:,t}$ is the embedding vector from the $t$-th column of $\*E_L$, $\*h_t \in \mathbb{R}^D$ is a hidden state, $\*c_t \in \mathbb{R}^D$ is a cell state.
$\mathsf{LSTM}_\mathsf{dec}()$ is another unidirectional LSTM layer with the initial hidden state $\*a_0 = \*h_{T}$ and the initial cell state $\*d_{0} = \*c_T$. The decoder LSTM receives zero vector $C$ times to produce the speaker-wise attractor vector $\*a_c$ for $C$ speakers.

EEND-EDA estimates the speaker label by comparing the embedding sequence $\*E_L$ with the speaker-wise attractors $\*A$:
\begin{align}
    \*Y = \mathsf{Sigmoid}(\*A^\top \*E_L). \label{eq:sigmoid}
\end{align}
As the number of attractor vectors can vary with the number of iterations $C$ in Eq.~\ref{eq:lstm_dec}, EDA can handle an unknown number of speakers by jointly estimating the number of iterations. In this paper, we fix the number of iterations to two since we only evaluate the method on two-speaker conversations.

With EEND-EDA described above, the speaker labels at all frames are estimated in parallel.
This parallel estimator lacks a mechanism handling the label dependency between frames.
In the next subsection, we consider the label dependency between frames by using intermediate speaker labels.

\subsection{Intermediate Attractors}
\label{sec:interattractor}

\begin{figure}[tb]
  \centering
  \includegraphics[width=\linewidth]{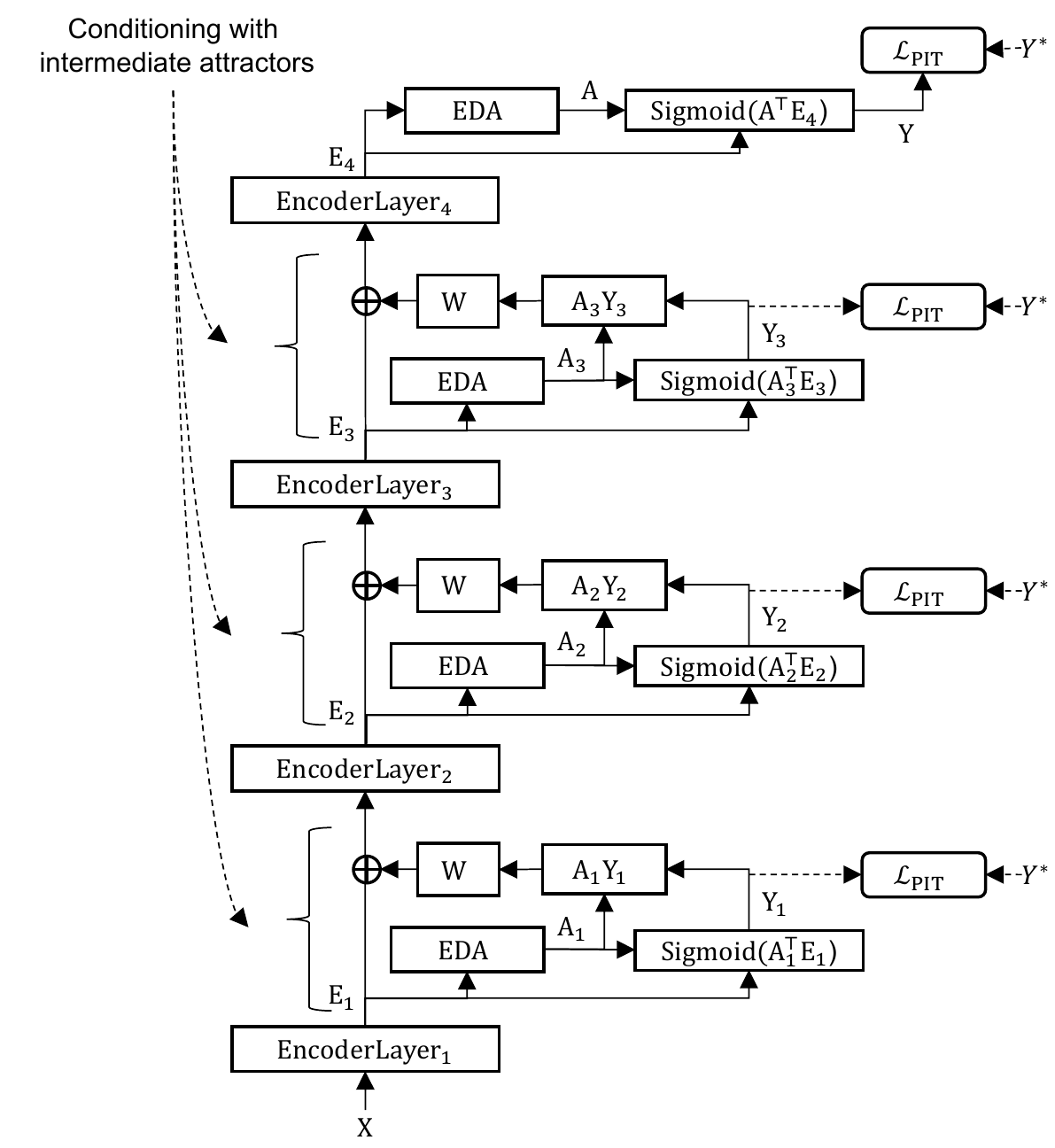}
  \caption{The overview of the proposed method with the four-layer EEND-EDA model.}
  \label{fig:interattractor}
\end{figure}

Considering the label dependency between frames, we extract ``intermediate'' speaker labels and feed them into the subsequent network layers.
Because the upper layer can refer to the whole sequence of intermediate speaker labels from the lower layer, conditional independence between frames is relaxed.
The overview of the proposed method with the four-layer EEND-EDA model is depicted in Fig. \ref{fig:interattractor}.

For producing the intermediate speaker labels, intermediate attractors are generated with EDA.
Using the same EDA components in Eq.~\ref{eq:eda}, intermediate attractors for the $l$-th layer are calculated as follows:
\begin{align}
  \*A_l = \mathsf{EDA}(\*E_l). \label{eq:intereda}
\end{align}
Then, intermediate speaker labels $\*Y_l$ for each layer are estimated using the intermediate attractors $\*A_l$, similar to Eq. \ref{eq:sigmoid}:
\begin{align}
    \*Y_l = \mathsf{Sigmoid}(\*A_l^\top \*E_l) \qquad (1 \le l \le L-1) \label{eq:intersigmoid}
\end{align}
The auxiliary permutation-invariant training loss is introduced for the intermediate speaker labels, and the model is trained with the summation of the original loss (Eq. \ref{eq:pit}) and the intermediate losses \footnote{We can take the weighted loss by introducing a mixing ratio. However, we ignore the hyperparameter in this work.}:
\begin{align} \label{eq:interloss}
    \mathcal{L}_\mathsf{inter} = \mathcal{L}_\mathsf{PIT}(\*Y^*, \*Y) + \frac{1}{L-1} \sum_{l=1}^{L-1} \mathcal{L}_\mathsf{PIT}(\*Y^*, \*Y_l).
\end{align}

For conditioning the subsequent network layers, the weighted intermediate attractors are added to the input embeddings.
Eq.~\ref{eq:enc} is modified by inserting the conditioning function for the input embedding:
\begin{align}
    \*E_{l} = \mathsf{EncoderLayer}_l(\mathsf{Condition}(\*E_{l-1})), \label{eq:cond}\\ 
   \mathsf{Condition}(\*E_l) = \*E_l + \*W \*A_l \*Y_l, \label{eq:cond2}
\end{align}
where $\*W \in \mathbb{R}^{D\times D}$ is learnable parameters that control the weights of intermediate predictions.$\*W$ is the only additional parameter introduced for the proposed method and is shared among $L-1$ layers.
$\*A_l \*Y_l$ in Eq.~\ref{eq:cond} can be interpreted as a $T$-length sequence of weighted averages of attractor vectors, and the weights are determined by the intermediate speaker labels. Through the intermediate attractors, the subsequent layers are conditioned on the intermediate speaker labels.

\subsection{Non-autoregressive Attractor Extraction}
\label{sec:nar}

\begin{figure}[t]
  \begin{center}
  \subfigure[Conventional EDA]{
    \includegraphics[width=0.75\linewidth]{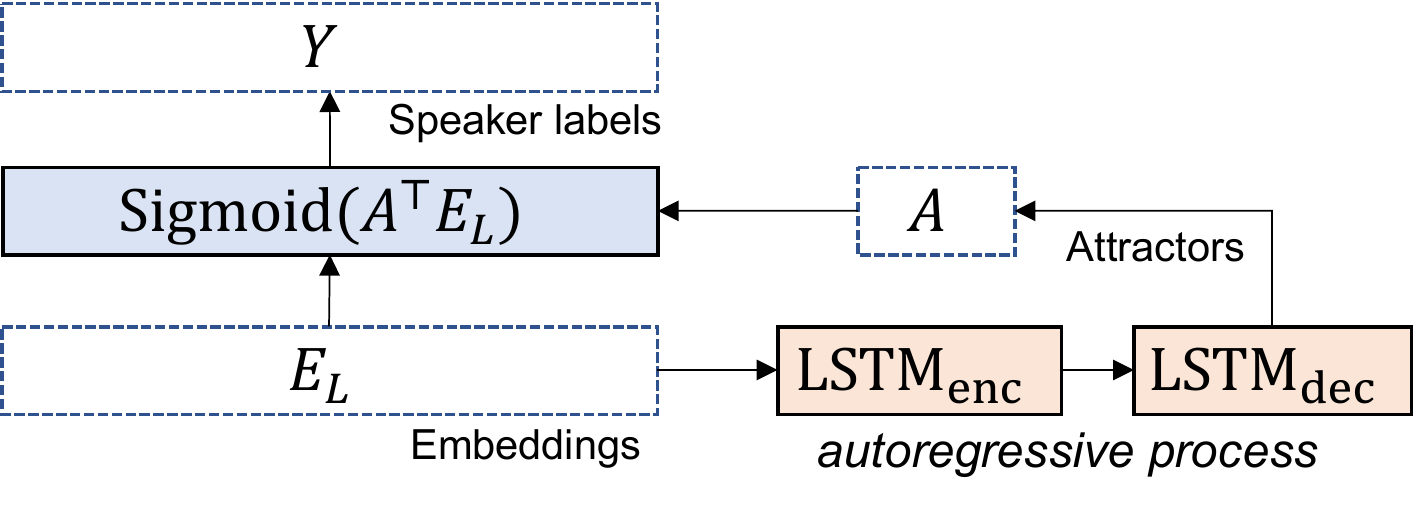}
    \label{fig:eda}
  }
  \subfigure[Proposed non-autoregressive attractor extraction]{
    \includegraphics[width=0.7\linewidth]{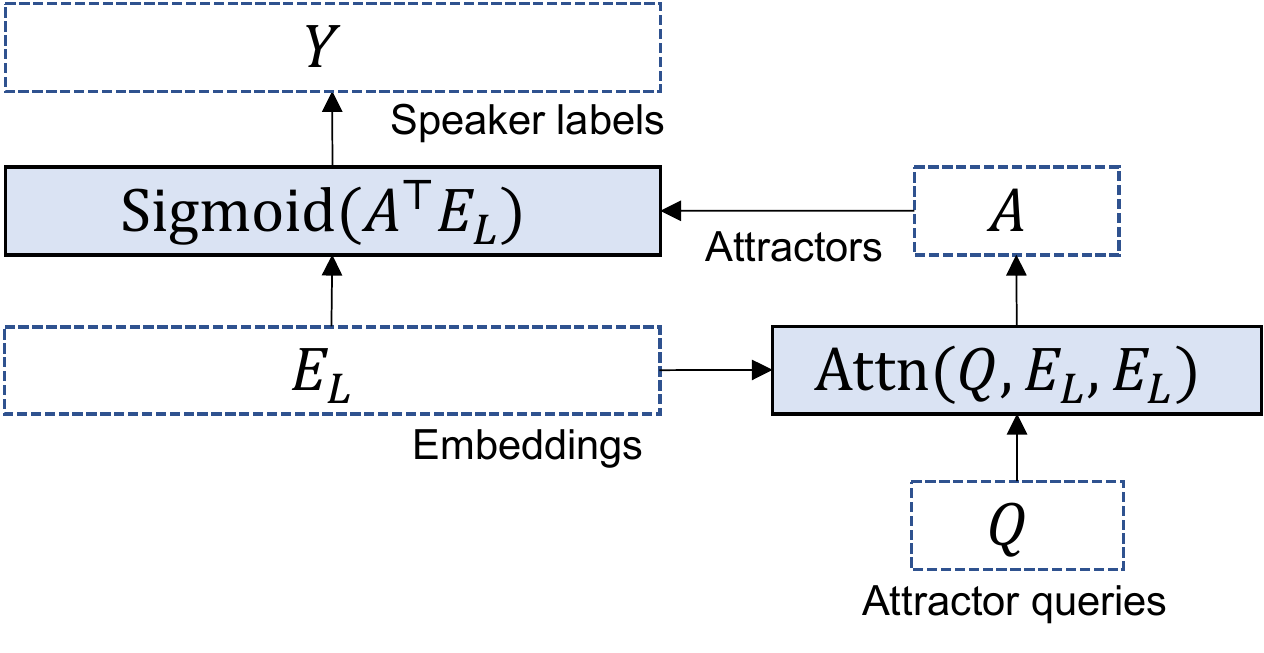}
    \label{fig:na}
  }
  \end{center}
  \caption{System diagrams of the conventional EEND-EDA and the proposed non-autoregressive attractor extraction.}
\label{fig:systems}
\end{figure}

For training efficiency with the intermediate speaker labels, we propose a non-autoregressive extraction of speaker-wise attractors instead of the LSTM-based autoregressive extractor used in EDA \cite{Horiguchi2020}. The difference is depicted in Fig.~\ref{fig:systems}.
We prepare query vectors $\*Q \in \mathbb{R}^{C\times D}$ for the attractors as learnable parameters through training.
A cross-attention module extracts the attractors by using the query vector and the frame-wise embeddings $\*E_L$ as keys and values:
\begin{align}
    \*A = \mathsf{Attn}(\*Q, \*E_L, \*E_L), \label{eq:crossattn}
\end{align}
where $\mathsf{Attn}$ is a multi-head attention layer used in Transformer decoders \cite{Vaswani2017}.
The intermediate attractors are extracted by using the same attention layer:
\begin{align}
    \*A_l = \mathsf{Attn}(\*Q, \*E_l, \*E_l) \quad (1 \le l \le L-1). \label{eq:intercrossattn}
\end{align}
Speaker labels and their intermediates are estimated using Eqs.~\ref{eq:sigmoid} and \ref{eq:intersigmoid}, respectively.
The intermediate labels are utilized to condition the subsequent layers using Eqs.\ref{eq:cond} and \ref{eq:cond2}.

Note that the original EDA can work with an unknown number of speakers, while our non-autoregressive extraction generates a fixed number of attractors.
Our method can be extended to an unknown number of speakers by adding a speaker-counting objective for the attractors, similar to EDA \cite{Horiguchi2020}.

\section{Experiments}
\label{sec:experiments}
\subsection{Data}

The statistics of the datasets are listed in Table \ref{tab:set}.
We followed the family of EEND works \cite{Fujita2019E2EDiarization, Fujita2019ASRU} to prepare test data for two-speaker telephone conversations extracted from CALLHOME \footnote{The data preparation code is available at \url{https://github.com/hitachi-speech/EEND}}.
We call CH-adapt for the adaptation dataset and CH-test for the test dataset.

Simulated two-speaker audio mixtures were used for a training set.
The source audio corpora were Switchboard-2 (PhaseI, II, III), Switchboard Cellular(Part1, 2), and the NIST Speaker Recognition Evaluation (2004, 2005, 2006, 2008). MUSAN corpus \cite{Snyder2015} was used for adding noise.
With a recently proposed mixture simulation algorithm \cite{landini22_interspeech}, simulated conversations (SimConv) were prepared using the statistics of the CH-adapt dataset.
Note that reverberation was not applied, which was the best configuration in \cite{landini22_interspeech}.

\begin{table}[tb]
\caption{Statistics of training/adaptation/test datasets.}
\label{tab:set}
\centering
\begin{tabular}{lrrr} \hline
 & Num. rec & Avg. dur & Overlap (\%) \\ \hline
SimConv & 24,179 & 368.8 & 8.1 \\
CH-adapt & 155 & 74.0 & 14.0 \\
CH-test & 148 & 72.1 & 13.0 \\ \hline
\end{tabular}
\end{table}

\begin{table*}[t]
\caption{Diarization error rates (\%) on CH-test, the number of parameters and training throughput (Tp; \#batches/sec) for each model.}
\label{tab:result}
\centering
\begin{tabular}{lrr|r|rrrrr|r|rrrrr} \hline
 & & &\multicolumn{6}{c|}{Before adaptation} & \multicolumn{6}{c}{After adaptation} \\
 & & & \multirow{2}{*}{DER} & \multicolumn{3}{c}{DER breakdown} & \multicolumn{2}{c|}{SAD} &
      \multirow{2}{*}{DER} & \multicolumn{3}{c}{DER breakdown} & \multicolumn{2}{c}{SAD} \\ 
 & \#Params & Tp &   & Miss & FA & CF & Miss & FA
   &   & Miss & FA & CF & Miss & FA  \\ \hline
EEND-EDA \cite{Horiguchi2020} & 6,402,305 & 3.30 & 8.66 & 3.7 & 4.2 & 0.7 & 0.8 & 2.0 & 7.74 & 5.0 & 2.2 & 0.5 & 2.0 & 0.6 \\
+InterLoss & 6,402,305 & 1.20 & 8.86 & 3.7 & 4.3 & 0.8 & 0.9 & 1.9 & 8.11 & 4.9 & 2.6 & 0.6 & 2.0 & 0.6 \\
+SelfCond & 6,468,097 & 1.03 & 9.29 & 3.9 & 4.6 & 0.8 & 0.8 & 2.2 & 9.13 & 3.9 & 4.3 & 1.0 & 0.8 & 1.9 \\
EEND-EDA-deep & 11,662,593 & 2.18 & 8.58 & 3.0 & 5.2 & 0.4 & 0.6 & 2.3 & 7.15 & 4.9 & 2.0 & 0.3 & 2.0 & 0.5 \\ \hline
Proposed EEND-NA & 5,613,056 & 4.15 & 11.15 & 6.1 & 2.8 & 2.2 & 1.6 & 1.4 & 11.34 & 6.8 & 2.9 & 1.6 & 2.7 & 0.5 \\
+InterLoss & 5,613,056 & 3.88 & 9.33 & 4.0 & 4.4 & 1.0 & 1.0 & 1.6 & 8.23 & 5.2 & 2.5 & 0.5 & 2.2 & 0.5 \\
+SelfCond & 5,678,848 & 3.79 & 8.81 & 3.9 & 4.3 & 0.6 & 1.1 & 1.9 & 7.77 & 5.0 & 2.5 & 0.3 & 2.1 & 0.6 \\
EEND-NA-deep & 10,873,344 & 3.26 & 9.51 & 5.0 & 3.1 & 1.4 & 1.2 & 1.5 & 9.45 & 5.9 & 2.7 & 0.9 & 2.2 & 0.6 \\
+InterLoss & 10,873,344 & 2.68 & 8.55 & 3.3 & 4.8 & 0.4 & 0.7 & 2.0 & 7.34 & 4.8 & 2.2 & 0.3 & 1.9 & 0.6 \\
+SelfCond & 10,939,136 & 2.50 & \bf 8.52 & 3.5 & 4.5 & 0.6 & 1.1 & 1.6 & \bf 7.12 & 4.7 & 2.1 & 0.3 & 2.1 & 0.5 \\
 \hline
\end{tabular}
\end{table*}

\begin{table}[tb]
\caption{Diarization error rates (\%) with intermediate speaker labels. The results are obtained using the EEND-NA-deep+SelfCond model after adaptation.}
\label{tab:result_inter}
\centering
\begin{tabular}{rrrrrrr} \hline
 \multirow{2}{*}{Layer}  & \multirow{2}{*}{DER}    & \multicolumn{3}{c}{DER breakdown} & \multicolumn{2}{c}{SAD} \\ 
  &   & Miss & FA & CF & Miss & FA \\ \hline
1 & 26.05 & 8.0 & 13.1 & 4.9 & 4.2 & 0.5 \\
2 & 12.76 & 6.6 & 4.0 & 2.2 & 3.2 & 0.6 \\
3 & 8.93 & 5.7 & 2.3 & 1.0 & 2.5 & 0.6 \\
4 & 7.89 & 5.2 & 2.2 & 0.6 & 2.1 & 0.6 \\
5 & 7.17 & 4.6 & 2.2 & 0.4 & 1.8 & 0.6 \\
6 & 7.11 & 4.6 & 2.1 & 0.3 & 2.0 & 0.6 \\
7 & \bf 7.01 & 4.6 & 2.1 & 0.3 & 2.0 & 0.5 \\
8(Last) & 7.12 & 4.7 & 2.1 & 0.3 & 2.1 & 0.5 \\
\hline
\end{tabular}
\end{table}

\subsection{Model Configurations}

\noindent\textbf{EEND-EDA:}
We used PyTorch-based implementation of EEND-EDA \footnote{\url{https://github.com/BUTSpeechFIT/EEND}} used in \cite{landini22_interspeech}.
Our baseline EEND-EDA model was exactly the same configuration as described in \cite{landini22_interspeech}, which is also the same as the original EEND-EDA paper \cite{Horiguchi2020}.
Audio features were 23-dimensional log-scale Mel-filterbanks computed every 10 msec.
For input to the neural network, 15 consecutive audio features were stacked to form 345-dimensional vectors every 100 msec.
We used four Transformer encoder layers with 256 attention units containing four heads.
Training configuration was also the same as \cite{landini22_interspeech}.
The training batch was a collection of 50-sec segments in the training set.
The batch size was 32.
Adam optimizer was used with the Noam learning scheduler and 200k linear warm-up steps.
The training was run for 100 epochs with the training set.
For adaptation, we run 100 epochs using the adaptation set with a learning rate of $10^{-5}$.
After training, the last 10 model checkpoints were averaged.

\noindent\textbf{EEND-EDA+InterLoss:}
The model is built with additional computations (Eqs.~\ref{eq:intereda}-\ref{eq:interloss}) introducing intermediate losses to the EEND-EDA model as described in Section~\ref{sec:interattractor}.
{\bf EEND-EDA+SelfCond} is built with additional computations  (Eqs.~\ref{eq:cond}-\ref{eq:cond2}) to apply the self-conditioning technique with the intermediate attractors.

\noindent\textbf{EEND-NA:}
We built a model named EEND-NA as EEND with the proposed non-autoregressive attractor extraction as described in Sec.~\ref{sec:nar}.
An +InterLoss model and a +SelfCond model were built on top of the EEND-NA model, similar to the EEND-EDA models.

\noindent\textbf{EEND-EDA-deep} and \textbf{EEND-NA-deep}:
We investigated the effect of deeper networks by increasing the number of Transformer blocks from four to eight.

\subsection{Metrics}

The diarization error rates (DERs) were evaluated with a forgiveness collar of 0.25s.
We also showed DER breakdown into miss, false alarm (FA), and confusion (CF) error rates, followed by speech activity detection (SAD) miss and false alarm errors.
The number of parameters (\#Params) and training throughput (Tp) for each model were shown to discuss the results with training efficiency. The training throughput was calculated on a Tesla V100 32G GPU.

\subsection{Results}

Table \ref{tab:result} shows the DERs on CH-test.
The first three rows show that the conventional EEND-EDA model could not benefit from the proposed intermediate labels and the self-conditioning technique.
Training throughput was down to one-third.
The results indicate that the LSTM-based autoregressive attractors cannot optimize the intermediate frame-level embeddings.

On the other hand, the proposed EEND-NA model showed performance improvement with the intermediate labels and the self-conditioning technique.
The training throughput of the proposed model was higher than that of the EEND-EDA model, and the slowdown by introducing the intermediate labels and self-conditioning was in an acceptable range.
The EEND-NA+SelfCond model reached a similar performance to the EEND-EDA model, while the proposed model has less number of parameters and higher throughput than the EEND-EDA model.
The results suggest that the non-autoregressive attractors can help optimize the intermediate frame-level embeddings, unlike EDA.
The EEND-NA model itself was worse than EEND-EDA, although the intermediate labels reduce the difference.
The disadvantage of the non-autoregressive attractors may come from the conditional independence between \textit{speakers}.
Using conditional inference on previously estimated speakers, like in the decoder part of EDA, may improve performance.

Table~\ref{tab:result} also shows the results of deeper (eight-layer) models.
The proposed EEND-NA-deep+SelfCond model achieved the best performance among the evaluated models.
While the conventional EEND-EDA-deep model achieved better performance than the original four-layer EEND-EDA model, we could not get the results with intermediate labels because of the slow training.
We expect that the EEND-EDA-deep model cannot benefit from the intermediate labels as with the baseline EEND-EDA models.

\subsection{Results with Intermediate Speaker Labels}

Table~\ref{tab:result_inter} shows diarization error rates with intermediate speaker labels produced by the EEND-NA-deep+SelfCond model.
The errors were consistently reduced layer by layer.
The results indicate that optimizing the speaker label at the lower layers improves the final diarization performance.
Unexpectedly, the seventh-layer results were better than the last eighth-layer results.
We think that conditioning with the intermediate labels could be enhanced by selecting the set of intermediate layers: not all the layers.

\section{Conclusion}
\label{sec:conclusion}
We proposed an end-to-end diarization model that uses intermediate speaker labels to condition the subsequent network layers.
The experiments showed that the intermediate labels with the non-autoregressive intermediate attractors boosted the diarization performance.
The proposed method with the deeper network benefits more from the intermediate labels, resulting in better performance and training throughput than EEND-EDA.

\bibliographystyle{IEEEbib}
\bibliography{refs}

\end{document}